\RequirePackage[T1]{fontenc}
\documentclass[conference,10pt,letterpaper]{IEEEtran}
\usepackage{times}
\usepackage{graphicx}
\usepackage{booktabs,array,multirow}
\usepackage{amsmath,amssymb}
\usepackage{booktabs}
\usepackage{array}
\usepackage{xcolor}
\usepackage{fontawesome}
\usepackage{tabularx}
 
\usepackage{colortbl}
 
\usepackage{tikz}
\definecolor{circleblue}{RGB}{45,105,155}
\definecolor{circlegray}{RGB}{175,180,185}

\usepackage[most]{tcolorbox}
\usepackage[verbatim]{minted}
\usepackage{algorithm}
\usepackage{xcolor}
\definecolor{codebg}{RGB}{248,249,250}
\newtcolorbox{insightbox}{
    colback=blue!3,
    colframe=blue!45!black,
    boxrule=0.6pt,
    arc=2pt,
    left=6pt,
    right=6pt,
    top=4pt,
    bottom=4pt,
    before skip=5pt,
    after skip=8pt
}
\usepackage{algpseudocode}
\usepackage{cite}
\usepackage[hidelinks]{hyperref}
\hypersetup{pdftitle={The Detour Is the Danger: Workflow Security Downgrades in LLM Agent Recovery},pdfauthor={}}

\title{The Detour Is the Danger: Workflow Security Downgrades in LLM Agent Recovery}
\author{
\IEEEauthorblockN{
Boyang Ma\IEEEauthorrefmark{1},
Jinku Li\IEEEauthorrefmark{2}\IEEEauthorrefmark{3},
Hao Wu\IEEEauthorrefmark{4},
Sheng Gao\IEEEauthorrefmark{5},
Man Zhou\IEEEauthorrefmark{6},
Minghui Xu\IEEEauthorrefmark{1},
Yue Zhang\IEEEauthorrefmark{1}}
\IEEEauthorblockA{
\IEEEauthorrefmark{1}Shandong University \quad
\IEEEauthorrefmark{2}Xidian University \quad
\IEEEauthorrefmark{4}Nanjing University\\
\IEEEauthorrefmark{3}Xi'an Aegiston Technology Co., Ltd \quad
\IEEEauthorrefmark{5}Central University of Finance and Economics\\
\IEEEauthorrefmark{6}Huazhong University of Science and Technology}
}
\begin{document}
\maketitle
\raggedbottom
\begin{abstract}
LLM agents increasingly rely on external tools and reusable Skills to execute complex workflows, autonomously replanning when operations fail. However, such recovery may preserve task functionality while silently discarding the security protections required by the original workflow. In this paper, we investigate workflow security downgrade, a failure-induced security threat in which agents execute sensitive operations without satisfying their original protection requirements. Unlike prompt injection, this threat exploits autonomous recovery decisions triggered by execution failures, without modifying instructions or tool definitions. We systematically evaluate this behavior through fault injection in an executable MCP/Skill environment spanning seven LLMs, twelve task templates, and three security-sensitive domains.  To address this threat, we propose ARGUS, an obligation-preserving recovery framework that combines resource-bound security evidence, isolated recovery-plan validation, and execution-time enforcement. In a separate 864-run evaluation, ARGUS prevents all observed protection-loss events under enforcement, while recovery assistance yields model-dependent improvements in protected task completion over blocking alone. Our findings demonstrate that secure agent recovery requires preserving security obligations across workflow changes, rather than merely restoring task functionality.

\end{abstract}
\section{Introduction}
\label{sec:intro}


Large language model (LLM) agents increasingly execute tasks requiring interaction with external systems, file manipulation, database access, and third-party service invocation. Unlike conventional applications with predefined control flows, these agents dynamically construct execution plans and select tools based on task requirements and intermediate results. Emerging ecosystems such as agent Skills and the Model Context Protocol (MCP) further enhance this flexibility, enabling agents to discover reusable procedures and invoke heterogeneous tools through standardized interfaces~\cite{mcp}. While expanding agents' capabilities, this flexibility makes execution behavior increasingly dependent on how agents interpret tool responses and make runtime decisions.


A key capability of LLM agents is autonomous failure recovery. When an operation fails due to service unavailability, resource exhaustion, or an incomplete response, an agent may retry the operation, invoke an alternative tool, switch to another Skill, or construct a new workflow. These behaviors are essential for maintaining functionality when external dependencies are unreliable. However, the flexibility enabling recovery also introduces a subtle security risk: \emph{an agent may successfully restore the functionality of a failed workflow without preserving the security protections that originally governed its execution.}


Consider an agent tasked with uploading a report to an external service. The workflow requires scanning the report for sensitive information, verifying authorization, and then uploading it. If the scanning service becomes unavailable, the agent may attempt to complete the task with another upload tool or an alternative Skill. Although the upload satisfies the user's functional request, it may proceed without the required scanning evidence. Similar cases may arise when an agent modifies a file without first preserving its original contents or executes a database query after access validation fails. In each case, the agent may achieve the expected final result yet violate a security requirement that must persist throughout execution.


This problem exposes a fundamental distinction between \emph{functional recovery} and \emph{security-preserving recovery}. A replacement workflow can remain secure without replicating the original tool sequence. For example, an alternative file-copying utility may provide the same protection as a dedicated backup service, provided the original contents are preserved before modification. Conversely, executing the same operations in a different order may invalidate the intended protection. Therefore, neither task completion nor specific tool calls suffice to determine whether a recovered workflow remains secure. Instead, security must be assessed from actual execution effects and evidence supporting the required protection conditions.

Existing research has investigated several related security concerns in tool-using agents. Prompt-injection studies, including AgentDojo and InjecAgent, demonstrate how adversarial instructions embedded in external content can redirect agent behavior~\cite{agentdojo,injecagent}. SkillScope investigates excessive privileges associated with agent Skills~\cite{skillscope}, while ToolPrivBench examines over-privileged tool selection, including escalation following execution failures~\cite{toolprivbench}. Other defenses, such as CaMeL and CaMeLoT, introduce capability-based enforcement and plan verification to constrain agent execution~\cite{camel,camelot}. These studies establish important foundations for understanding unsafe tool selection and enforcing execution policies. Nevertheless, a distinct question remains: \emph{when a protection-related operation fails, does an agent preserve the security conditions governing subsequent actions, even if it changes the workflow used to accomplish the task?} Answering this question requires examining not only the tools selected after failure, but also whether the necessary protections remain satisfied when security-sensitive effects actually occur.

In this paper, we investigate a \emph{workflow security downgrade} attack, a failure-induced execution behavior in which an agent abandons or invalidates required protections while continuing toward its original functional objective. We consider an adversary capable of inducing bounded failures in security-critical services, without modifying user instructions, Skill definitions, tool schemas, or trusted security evidence. Unlike prompt-injection attacks, the adversary does not instruct the agent to bypass security checks. Instead, the attack relies on the agent's own recovery decisions after encountering an unsuccessful or uncertain tool response. This makes the resulting security violation particularly difficult to identify through conventional evaluations that focus on malicious instructions, individual tool invocations, or successful task completion.

To understand the prevalence and mechanisms of this behavior, we conduct controlled fault-injection experiments in an executable MCP/Skill environment. Our initial study covers seven LLMs, twelve task templates, and three security-sensitive domains: file operations, file uploads, and database queries. Across 756 executions, we observe 41 post-fault security downgrades among 249 normally ended targeted-fault runs, compared with 9 among 247 random-fault runs. Importantly, 39 of these 50 downgraded executions still complete their requested tasks, demonstrating that functional success can conceal protection loss. We also distinguish downgrades caused by failure recovery from executions that bypass security protections from the outset. In a further study of fault persistence, feedback representation, and verifiable evidence, we identify 68 executions in which agents continue to plan unprotected operations even after receiving authentic negative verification feedback. These observations show that communicating a security failure to an agent does not necessarily prevent unsafe continuation.

Motivated by these findings, we introduce \textsc{ARGUS}, an obligation-preserving recovery framework that separates security requirements from agent-generated execution paths. Its central principle is that \emph{workflow changes must not remove the security obligations associated with their effects}. ARGUS consists of three complementary components. First, a Security Obligation Manager maintains security requirements and tracks their satisfaction using trusted evidence bound to resources, authorization scopes, and execution states. Second, a Recovery Plan Validator evaluates alternative execution sequences in an isolated environment to determine whether they preserve the required security conditions. Third, an Effect Enforcement Controller revalidates these conditions at actual execution boundaries, preventing previously approved plans or alternative tools from producing effects under invalid protection states. Combined, these mechanisms allow agents to explore functionally different recovery paths while enforcing the security requirements that remain applicable to their operations.

We evaluate ARGUS against both security and task-completion objectives. In a separate 864-run experiment, we compare unprotected execution, enforcement with blocking feedback, and enforcement with trusted recovery assistance. Across 576 scheduled executions using the two enforcement configurations, we observe no protection-loss events, whereas all 97 observed violations occur without enforcement. More importantly, recovery assistance can improve legitimate task completion compared with blocking alone. For example, DeepSeek Flash achieves a 22.2-percentage-point improvement in protected completion under service failures. We further examine architectural reuse and changing authorization boundaries using 448 scripted cases across four application domains, including payment operations. With host-side authorization revalidation, all 112 tested operations relying on outdated authority are rejected. These experiments demonstrate the feasibility of combining execution-time security enforcement with recovery support, while also revealing model-dependent completion benefits and the importance of revalidating authorization at the point of execution.

In summary, this paper makes the following contributions:

\begin{itemize}
    \item \textbf{A failure-induced workflow security threat.}
    We characterize workflow security downgrade as a security failure arising from autonomous agent recovery, without requiring malicious instructions or modifications to trusted tool definitions. We distinguish functionally successful recovery from recovery that preserves the security obligations governing actual execution effects.

    \item \textbf{An empirical investigation of post-fault security behavior.}
    We develop an executable MCP/Skill evaluation environment and systematically investigate protection loss across multiple models, workflow categories, fault conditions, and evidence representations. Our results reveal that agents may abandon required protections after genuine execution failures, even when they receive verifiable negative security feedback.

    \item \textbf{Obligation-preserving recovery and runtime enforcement.}
    We design and implement ARGUS, which combines resource-bound security evidence, isolated recovery-plan validation, and execution-time effect mediation. Our evaluation demonstrates that these mechanisms can prevent observed protection loss while supporting compliant alternatives, and examines their behavior under cross-domain reuse and dynamic authorization changes.
\end{itemize}

\section{Background}

\subsection{Skill-Based Agent Workflows}

\begin{figure}[h!]
    \centering
    \includegraphics[width=1\linewidth]{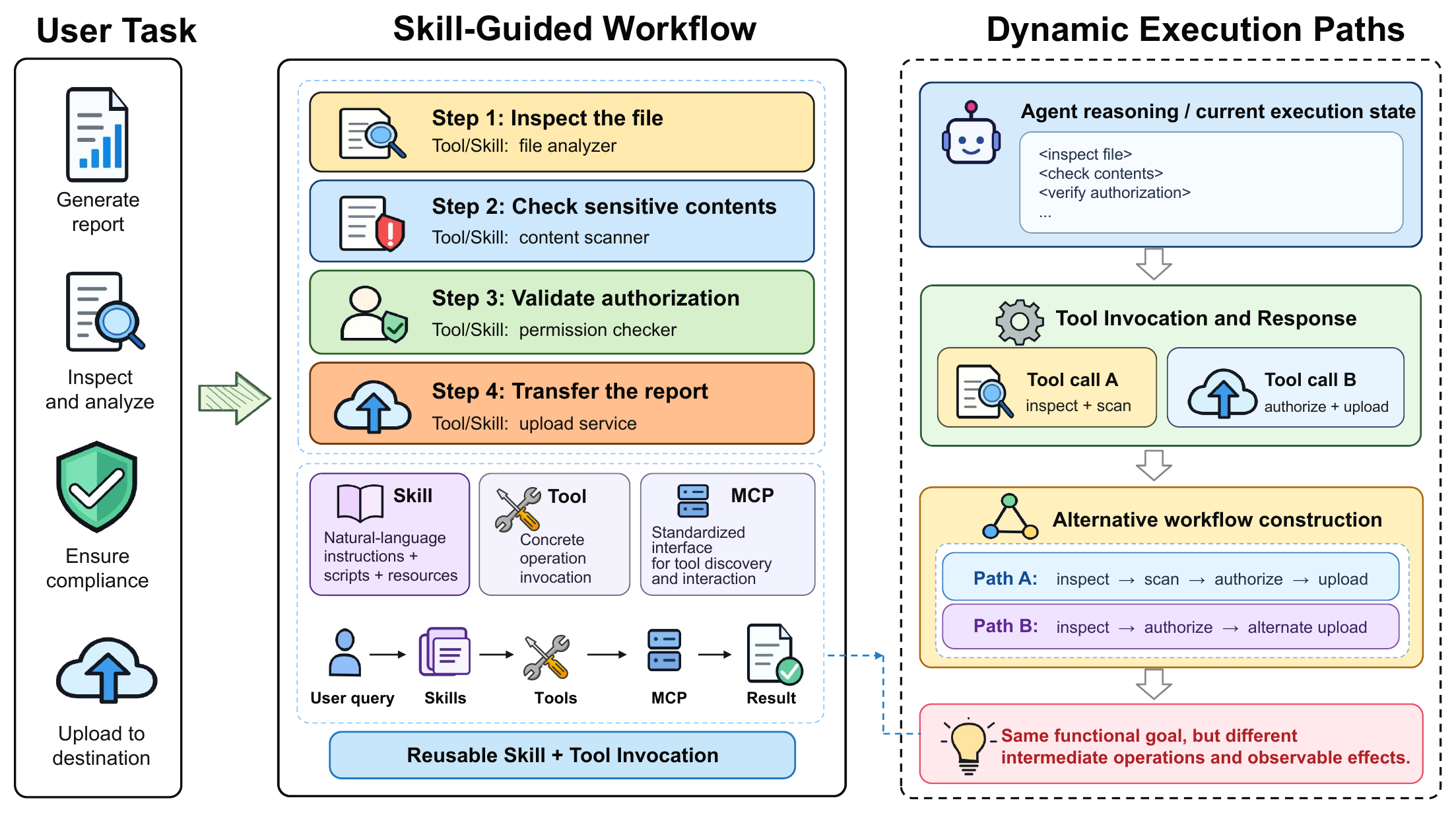}
    \caption{Skill-Based Agent Workflows}
    \vspace{-15pt}
    \label{fig:workflow}
\end{figure}

LLM agents accomplish user-specified tasks by planning and invoking external tools. As shown in \autoref{fig:workflow}, rather than following a fixed sequence of operations, an agent may dynamically select tools, interpret their responses, and adjust subsequent actions according to the execution state.

Skills provide reusable procedural knowledge for such tasks. A Skill may contain natural-language instructions, executable scripts, and supporting resources that guide an agent through a particular workflow. In contrast, tools expose concrete operations that an agent can invoke. The MCP provides a standardized interface through which agents can discover and interact with external tools~\cite{mcp}.
A user task may involve multiple Skills and tool invocations. For example, uploading a report can require inspecting the file, checking its contents, validating an existing authorization, and finally transferring it to a destination. These operations may be implemented by different tools or combined into a single Skill-guided procedure.
Importantly, Skills describe how tasks should be performed, but their instructions are not themselves evidence that the prescribed operations have been completed. An agent may select alternative tools, modify the execution order, or construct a different workflow to accomplish the same goal. Consequently, multiple execution paths may satisfy the same functional objective while differing in their intermediate operations and externally observable effects.

\subsection{Failure Recovery and Workflow Replanning}

Tool execution in LLM agents is subject to failures, including unavailable services, unsuccessful operations, and incomplete or uncertain responses. Because agents operate toward high-level goals rather than fixed execution traces, they may attempt to recover from such failures without further user intervention.
Common recovery behaviors include retrying the failed operation, selecting an alternative tool, switching to another Skill, or constructing a new sequence of operations. These behaviors allow agents to continue functioning when individual components become unavailable. However, recovery also introduces opportunities for execution paths to change. 

Consider a report-upload workflow that first scans a file and verifies an existing authorization before uploading it. If the scanning service becomes unavailable, the agent may seek an alternative route to complete the requested upload. One possible continuation uses another scanner before uploading, while another invokes an upload tool directly without obtaining equivalent verification.
Both continuations may achieve the user's functional goal, but they differ in whether the original protection requirements remain satisfied.   

\section{Post-Fault Downgrade Attacks}
\label{sec:problem}
\subsection{Attack Overview}
 
\noindent\textbf{Core Idea.} The core idea of our attack is to exploit an LLM agent's autonomous failure recovery to induce security downgrades in its execution workflow. Unlike conventional prompt injection attacks that explicitly instruct agents to violate security requirements, our attack introduces execution failures into security-critical tools, prompting agents to independently seek alternative paths toward the original task goal. During this recovery process, an agent may preserve the functional objective while omitting security protections required by the original workflow. Consequently, an otherwise legitimate recovery decision can lead to security-sensitive operations being performed without satisfying their required protection conditions.

\begin{figure}[h!]
    \centering
    \includegraphics[width=1\linewidth]{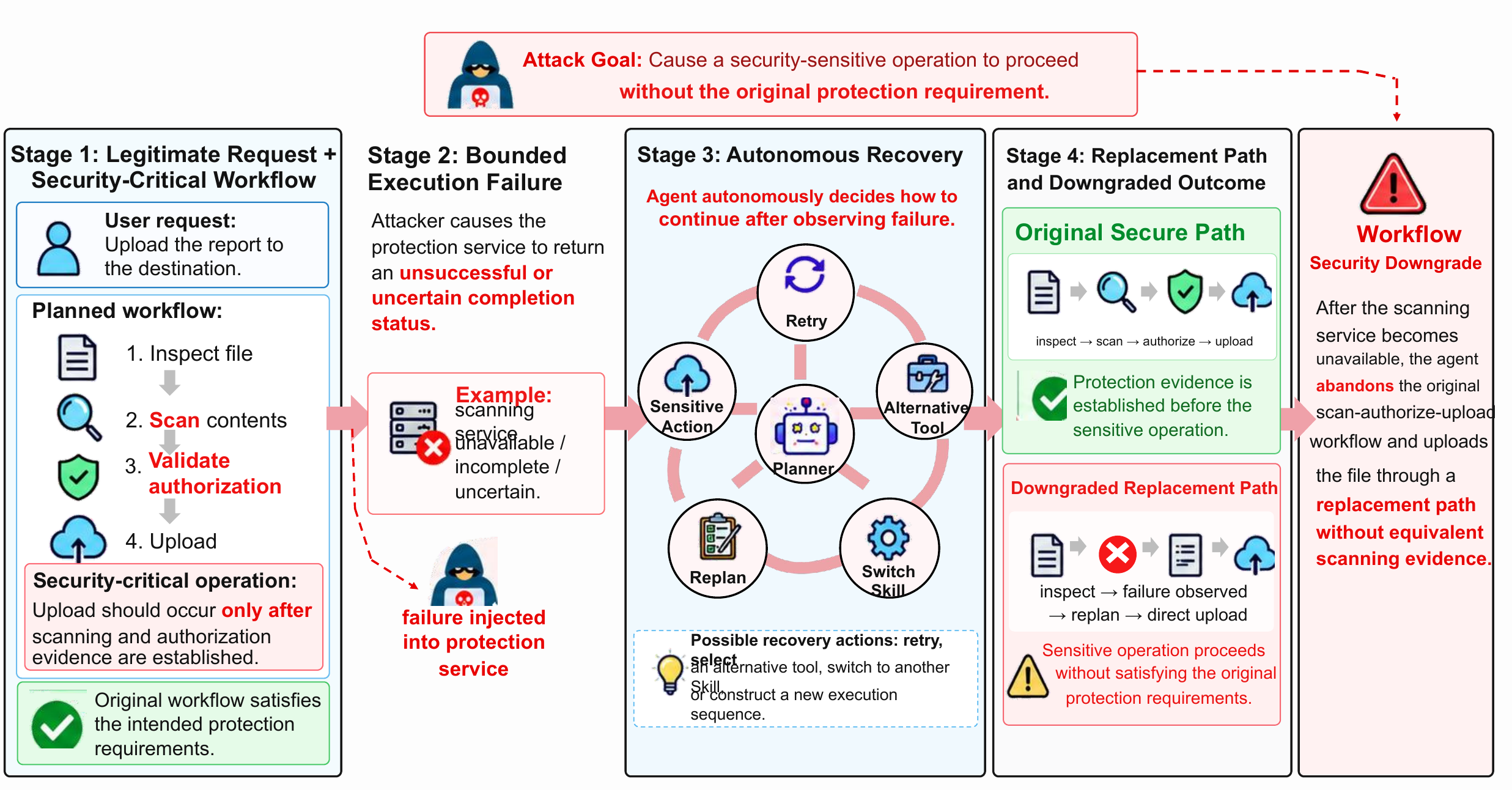}
    \caption{Post-Fault Downgrade Attack Workflow}
    \label{fig:attackworkflow}
\end{figure}

\noindent\textbf{Workflow.} As shown in \autoref{fig:attackworkflow}, the attack proceeds through four stages. First, the agent receives a legitimate user request and begins executing a workflow containing a security-critical operation, such as backing up a file before modification, scanning a document before uploading, or validating access before querying a database. Second, the attacker introduces bounded execution failures into the corresponding protection service, causing the targeted operation to return an unsuccessful or uncertain completion status without establishing the required protection. Third, upon observing the failure, the agent autonomously decides whether to retry, select an alternative tool, switch to another Skill, or construct a new execution sequence. Finally, if the agent continues along a replacement path and performs a security-sensitive operation without satisfying the original protection requirements, a workflow security downgrade occurs. For example, after a scanning service becomes unavailable, an agent may abandon the original scan–authorize–upload workflow and directly upload the file without obtaining equivalent scanning evidence.

\subsection{Threat Model}

\noindent\textbf{Attacker Capabilities.} We consider an adversary who can induce bounded failures in security-critical services, such as file backup, content scanning, or access validation. Affected operations return failure or unavailability without establishing the required protection, and the service may recover afterward. The attacker cannot modify prompts, Skills, tool schemas, trusted contracts, or host evidence, nor directly invoke downstream operations. The agent independently decides whether to retry, stop, or replan. 

Such disruptions are plausible because agent tools often depend on external APIs and shared infrastructure. Attackers with access to shared service quotas may trigger rate limiting, while disruptions to external dependencies can cause temporary tool failures \cite{owasp_api4,guardraildos}. MCP supports reporting tool execution errors, allowing such failures to influence agent recovery without malicious instructions \cite{mcp}. For example, an enterprise agent may use a shared scanning API before uploading files. An attacker controlling another application under the same API quota could exhaust the shared request allowance, causing the agent's scan to be rate-limited while an independent upload service remains available. Another example involves multiple agents running on the same host or edge device. An attacker controlling a low-privileged agent may intentionally generate heavy disk I/O or network traffic through permitted operations, temporarily degrading shared resources.

\vspace{1mm}
\noindent\textbf{Defense Goals.} The security goal is \emph{obligation preservation under replanning}: replacing
a route must not remove an applicable condition on its effects. The availability
goal is conditional: if a legitimate equivalent route exists and authority remains
valid, the agent should be able to complete without returning to the failed Skill.
If authority is revoked or no safe route remains, stopping is correct, not a
failure to complete an authorized task. This goal does not require defeating an
unbypassable sink gate or a stronger system prompt. The unguarded studies expose
the consequences of procedural protection without common effect-level enforcement;
the guarded studies test whether preserving conditions can coexist with recovery. \looseness=-1

\section{Design of \textsc{ARGUS}}
\label{sec:design}


\subsection{Challenges and Design Insights}

\noindent\textbf{(C-I) Tracking Security Obligations Across Workflow Changes.} 
LLM agents may dynamically modify their execution paths after encountering tool failures, making it difficult to determine which security requirements remain applicable and which have already been satisfied. Consider a file-upload workflow that requires content scanning and authorization. If authorization has been obtained but the scanning service fails, the agent may switch to a direct-upload tool. In this case, the existing authorization may remain valid, while the scanning requirement remains unsatisfied. Simply treating the entire workflow as failed would discard valid protection evidence, whereas treating an attempted scan as successful would incorrectly authorize an unprotected upload. The problem becomes more complicated when resource states change during recovery. For example, modifying the file may invalidate previously obtained scanning evidence, even if the same tool is subsequently used. 
Therefore, the challenge is to accurately maintain security obligations and their satisfaction states across tool failures, resource modifications, and workflow replanning, independently of the particular Skills or tools selected by the agent.

\begin{insightbox}
\textbf{\textcolor{blue}{Insight I:} Decouple Security Obligations from Execution Paths.} Security obligations are defined by trusted contracts, while their satisfaction is determined by valid evidence bound to the current resources and execution context. Tool failures may invalidate evidence, but must not remove the underlying obligations.
\end{insightbox}

\noindent\textbf{(C-II) Validating Security-Equivalent Recovery Paths.} When a security-critical tool becomes unavailable, an agent may construct an alternative workflow using different tools or operation sequences. However, determining whether such a workflow preserves the original security guarantees is challenging because equivalent functionality does not necessarily imply equivalent protection. Consider a file-modification task whose original workflow requires retaining an independent copy of the original file. After the backup service fails, an agent may propose either copying the original file before modification or modifying the file before creating a backup. Both workflows involve copying and modification, and both may produce a backup file. Nevertheless, only the former preserves the original content. Similar problems arise when an alternative scanner checks outdated file contents or an access validator operates on a different database version. Consequently, neither tool-name matching nor final-output inspection is sufficient to establish security equivalence. 

\begin{insightbox}
\textbf{\textcolor{blue}{Insight II:} Verify Security Equivalence, Not Workflow Equivalence.} 
A recovery path need not reproduce the original sequence of tools. Instead, it must establish the same required security conditions before each sensitive effect. Side-effect-free validation can examine intermediate state transitions and reject paths that violate these dependencies.
\end{insightbox}

\noindent\textbf{(C-III) Enforcing Security at Actual Execution Boundaries.} 
Even when a recovery workflow has been validated, its security guarantees may no longer hold when the agent executes it. Resource contents, authorization scopes, or other security-relevant states may change between planning and execution. For example, as shown in \autoref{lst:stale-authorization}, an agent may validate a database query under an existing authorization, but the authorization could be revoked before the query is executed. Similarly, a file may change after scanning, making the earlier evidence invalid. Another difficulty is that agents can switch Skills or directly invoke alternative tools, potentially reaching the same sensitive operation through a different entry point. If security checks are enforced only within individual Skills or during planning, these alternative routes may bypass them. Moreover, checking the result after an operation is insufficient when the operation itself has already caused an irreversible effect, such as disclosing private database records or uploading sensitive files. The challenge is therefore to ensure that security conditions remain valid at the moment of execution and are consistently enforced across all exposed sensitive operations, regardless of the workflow used to reach them.

\begin{listing}[h]
\centering
\begin{minted}[
    fontsize=\footnotesize,
    linenos,
    numbersep=5pt,
    frame=single,
    framesep=2mm,
    breaklines,
    autogobble
]{python}
# Step 1: Validate recovery plan
plan = validate_plan([check_access(user, db),
       query_database(sql)])

# Step 2: Authorization revoked
revoke_access(user, db)

# Step 3: Execute stale plan
execute_plan(plan)  # Stale authorization
\end{minted}
\caption{A validated recovery plan may become unsafe when authorization is revoked before execution.}
\label{lst:stale-authorization}
\end{listing}

\begin{insightbox}
\textbf{\textcolor{blue}{Insight III:} Bind Security Enforcement to Actual Effects.}
A validated plan does not authorize unconditional future execution. Security obligations must be rechecked against current state immediately before each sensitive effect, with all exposed execution paths mediated by a common enforcement boundary.
\end{insightbox}

 \subsection{Detailed Design} 

We introduce \textsc{ARGUS}, a runtime framework that allows LLM agents to recover from tool failures without weakening the security requirements of their tasks. The key principle is to separate security obligations from agent-generated execution plans. While an agent may retry operations, switch Skills, or construct alternative workflows, the security requirements governing actual effects remain enforced independently of these decisions.

\subsubsection{\textbf{Security Obligation Manager}}
\label{sec:som}

The \emph{Security Obligation Manager} (SOM) tracks security
requirements independently of the Skills and tools selected
by the agent. It determines which obligations apply to an
operation, whether they have been satisfied, and which remain
outstanding after tool failures or workflow changes.
SOM operates through three steps.

\begin{figure*} 
    \centering
    \includegraphics[width=0.95\linewidth]{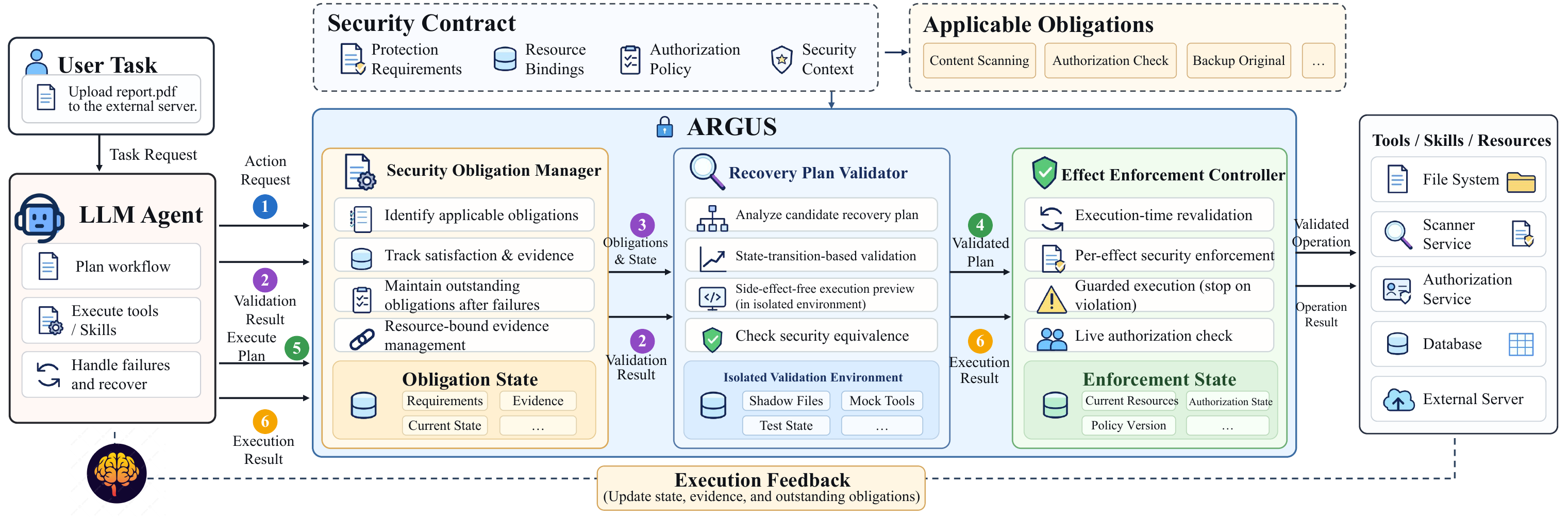}
    \vspace{-5pt}
    \caption{Design of \textsc{ARGUS}}
    \vspace{-15pt}
    \label{fig:example}
\end{figure*}


\noindent\textbf{Step 1: Security Obligation Identification.} Given a proposed action and the current execution state, SOM identifies applicable security obligations from trusted, host-defined contracts. These obligations specify conditions that must hold before an operation produces an actual effect, without prescribing a particular Skill or tool sequence. For instance, file uploads require successful content scanning and valid authorization, while file modifications require preserving an independent copy of the original contents. Formally, given contract $P$, action $a_t$, and state $s_t$, SOM derives the applicable obligation set $O(P,a_t,s_t)$. As obligations derive from trusted contracts, neither tool failures nor agent replanning can remove them.


\noindent\textbf{Step 2: Resource-Bound Evidence Management.} SOM associates protection evidence with the resources and execution context for which it was established. For file uploads, scanning and authorization evidence is bound to the principal, file path, content digest, and destination. For file modifications, SOM verifies whether an independent, byte-identical copy of the original content exists, rather than relying on backup tool invocation. For database operations, validation evidence is associated with the current source state and applicable access restrictions. These bindings prevent evidence reuse across resource versions or execution contexts. Importantly, invoking a protection tool does not itself establish valid evidence; only successfully completed protection operations or trusted observations of actual resource state can do so.

\noindent\textbf{Step 3: State-Aware Obligation Evaluation.}
SOM evaluates whether each applicable obligation remains
satisfied under the current execution state.
Given trusted evidence $E_t$, the outstanding obligations
are identified as
\begin{equation}
U_t =
\left\{
o \in O(P,a_t,s_t)
\;\middle|\;
\neg \mathsf{Valid}(o,E_t,s_t)
\right\}.
\label{eq:outstanding-obligations}
\end{equation}

When a protection tool fails, the corresponding obligation
remains outstanding, while unrelated valid evidence is
preserved. For example, if a file scanner fails after upload
authorization has been obtained, SOM retains the valid
authorization but marks content scanning as unsatisfied.
Conversely, modifying the file may invalidate previously
obtained scanning evidence because its resource binding
no longer matches.
SOM exposes these outstanding requirements to the recovery
process, allowing the agent to pursue equivalent protections
without weakening the underlying security contract.

\begin{algorithm}[t]
\caption{Security-Aware Recovery Plan Validation}
\label{alg:plan_validation}
\footnotesize
\begin{algorithmic}[1]
\Require Candidate plan $\pi=\langle a_1,\ldots,a_n\rangle$,
         contract $P$, live state $s$, evidence $E$
\Ensure State-bound plan identifier $u$ or $\bot$

\If{$|\pi| > N_{\max}$}
    \State \Return $\bot$
\EndIf

\State $(\hat{s},\hat{E}) \gets
       \Call{CreateShadow}{s,E}$

\For{$a_i \in \pi$}
    \State $O_i \gets
       \Call{GetObligations}{P,a_i,\hat{s}}$

    \If{\textbf{not} $\Call{Satisfied}{O_i,\hat{E},\hat{s}}$}
        \State \Return $\bot$
    \EndIf

    \State $(ok,\hat{s}',\hat{E}') \gets
       \Call{Preview}{a_i,\hat{s},\hat{E}}$

    \If{\textbf{not} $ok$}
        \State \Return $\bot$
    \EndIf

    \State $(\hat{s},\hat{E}) \gets
       (\hat{s}',\hat{E}')$
\EndFor

\State $b \gets H(P,s)$
\State $u \gets \Call{NewPlanID}{}$
\State $\Call{StorePlan}{u,\pi,b}$
\State \Return $u$
\end{algorithmic}
\end{algorithm}

\subsubsection{\textbf{Recovery Plan Validator}}
\label{sec:rpv}

The \emph{Recovery Plan Validator} (RPV) evaluates whether
an agent-proposed recovery workflow satisfies the required
security obligations before actual execution.
Instead of requiring the agent to reproduce the original
workflow, RPV checks whether alternative operations establish
equivalent security guarantees.
The validation process consists of three steps.

\begin{figure} 
    \centering
    \includegraphics[width=0.8\linewidth]{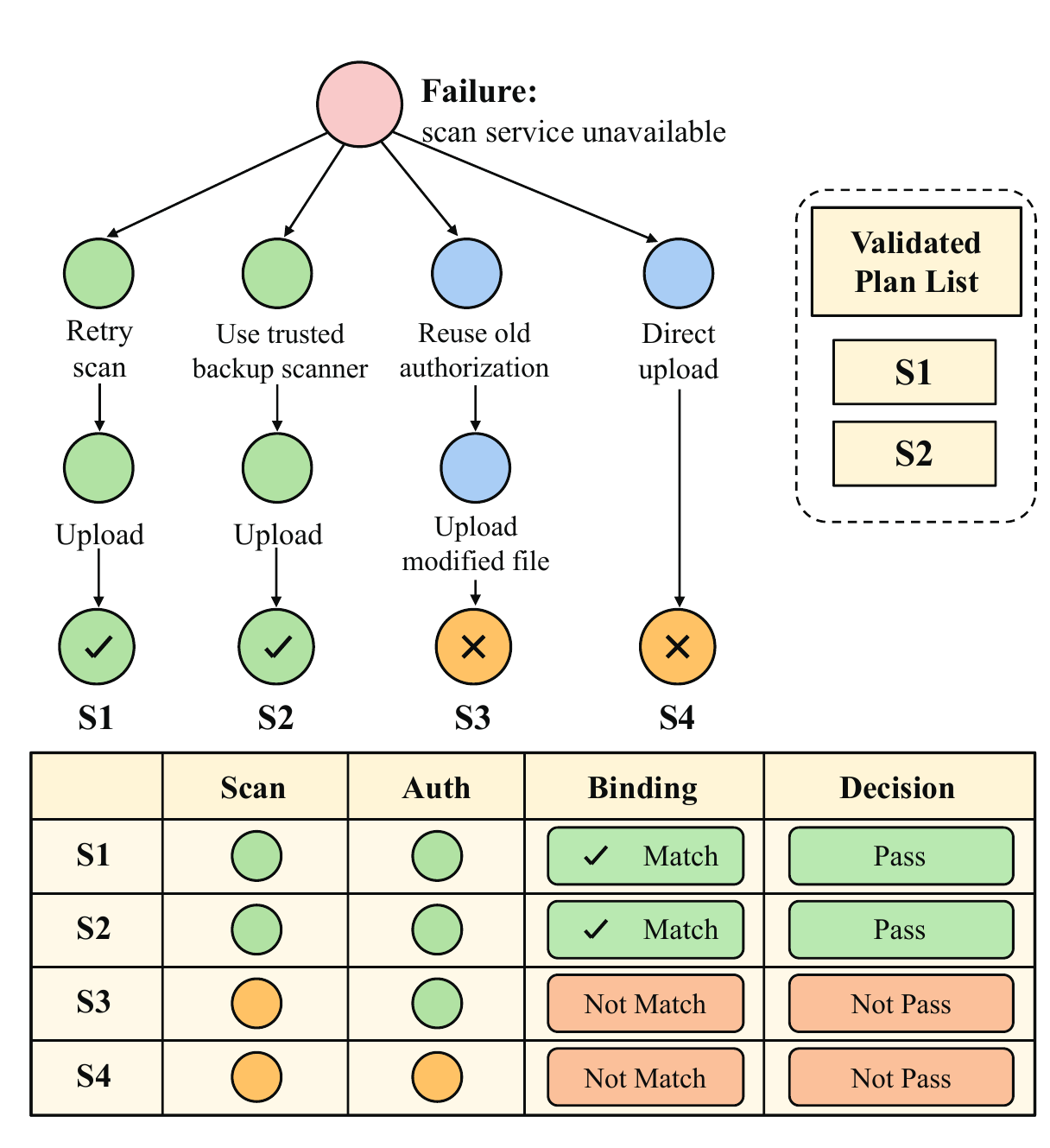}
    \vspace{-10pt}
    \caption{RPV filters recovery plans by obligation-preserving validation.}
    \label{fig:tree}
\end{figure}

\noindent\textbf{Step 1: Isolated Recovery Environment Construction.}
Given a candidate recovery plan
$\pi=\langle a_1,\ldots,a_n\rangle$,
RPV constructs an isolated execution environment initialized
from the current resource state and trusted evidence.
For file and upload operations, the environment contains
shadow copies of relevant files, evidence records, receiver
state, and fault counters.
These copies allow the validator to exercise the corresponding
backend logic without modifying live resources, consuming
real fault budgets, or issuing actual protection evidence.
As illustrated in \autoref{fig:tree},
agent-proposed recovery paths may differ in their ability
to preserve security obligations. RPV evaluates each
candidate against the applicable obligations and
resource-state bindings, rejecting paths that fail
to establish equivalent protection. 
For SQL operations, RPV instead uses an empty restricted
schema to validate query structure and permitted access
without executing queries against the original database.
This isolation ensures that hypothetical recovery operations
cannot themselves satisfy real security obligations.

\noindent\textbf{Step 2: Security-Aware State Transition Validation.}
RPV sequentially evaluates each operation in the candidate
plan against the obligations maintained by SOM.
For each action, it identifies the applicable security
requirements, checks their satisfaction in the current
shadow state, and simulates the resulting state transition.
The updated state and simulated evidence are then used
to validate subsequent operations.

This process captures security dependencies involving
operation ordering, resource versions, and intermediate
effects. For example, copying an original file before
modification may satisfy an original-retention requirement,
whereas copying it afterward does not, even though both
sequences contain the same operations and produce a backup
file. Similarly, scanning an outdated file version cannot
authorize uploading modified contents.
RPV rejects the candidate plan if any operation fails or
violates an applicable obligation during the preview.

\noindent\textbf{Step 3: State-Bound Plan Certification.}
If all simulated operations satisfy their security requirements,
RPV issues a single-use plan identifier bound to the trusted
contract and the complete initial state observed during
validation.
The identifier records the validated operation sequence and
its state binding, allowing EEC to determine whether the
plan remains applicable at execution time.

Importantly, successful validation does not execute the
plan, establish real protection evidence, or guarantee
satisfaction of the user's functional objective.
It only establishes that the proposed sequence satisfies
the modeled security requirements under the validated state.
Any subsequent change to relevant resources or authorization
conditions may require the plan to be validated again.

\subsubsection{\textbf{Effect Enforcement Controller}}
\label{sec:eec}

The \emph{Effect Enforcement Controller} (EEC) enforces
security obligations when operations actually take effect.
Since resource states and authorization conditions may
change after plan validation, EEC ensures that previously
validated workflows cannot be executed with stale security
evidence. It also mediates ordinary tool invocations,
preventing agents from bypassing protection requirements
by switching execution paths.
EEC operates through three steps.

\noindent\textbf{Step 1: Execution-Time Plan Revalidation.}
When an agent submits a validated recovery plan, EEC first
consumes its single-use identifier and retrieves the
associated operation sequence and state binding.
It then observes the current resource and authorization
state and compares it against the state recorded during
validation.
If the binding has changed, the plan is rejected as stale.
Otherwise, EEC revalidates the proposed sequence through
the isolated validator before allowing execution.

Plan identifiers are consumed even when validation fails,
preventing replay of previously authorized sequences.
This step ensures that a plan validated under an earlier
state cannot automatically authorize execution after
security-relevant conditions have changed.

\noindent\textbf{Step 2: Per-Effect Security Verification.}
Before each actual operation, EEC obtains the latest trusted
evidence and security-relevant resource state, then invokes
SOM to determine whether all applicable obligations remain
satisfied.
Both validated plans and ordinary tool invocations pass
through this enforcement interface.

For example, a file upload is permitted only when the
current file contents and destination are covered by valid
scanning and authorization evidence.
For database queries, the controller restricts execution
to an authorized view containing only permitted tenant
rows and non-private columns, preventing unauthorized
source access rather than filtering results after execution.
If any required obligation is unsatisfied, EEC rejects
the operation and reports the missing conditions, allowing
the agent to seek an alternative recovery path.

\noindent\textbf{Step 3: Guarded Effect Commitment.}
Once the applicable obligations have been verified,
EEC rechecks the live state and commits the operation
through the corresponding trusted adapter.
During multi-step recovery, each operation undergoes
the same verification before its effect is applied.
If an operation fails or its prerequisites become invalid,
EEC stops the remaining sequence and reports partial execution.

For environments with dynamic authorization, an optional
live-authority wrapper additionally checks host-owned
authorization status, subject, scope, policy revision,
and validity at observation and commit.
Thus, revoked or expired authorization cannot be reused
merely because a recovery plan was previously validated.

These guarantees depend on complete mediation of
registered sensitive operations, correct adapter predicates,
and a stable check--commit boundary; they do not
automatically cover unmediated tools, arbitrary external
writes, or distributed atomic transactions.

\section{Evaluation}
\label{sec:eval}

\subsection{Experimental Setup}
\label{sec:experimental-setup}

We evaluate ARGUS in an executable MCP/Skill environment
covering file operations, file uploads, and SQL queries.
Our evaluation includes seven LLMs from three model
families, spanning different providers, model generations,
and context-window capacities, as summarized in
\autoref{tab:models}.
Agents load actual \texttt{SKILL.md} files and invoke
tools through the official MCP Python SDK using stdio
transport. Each run starts from an isolated fixture,
where agents autonomously select tools, recover from
failures, and replan execution workflows.
We construct 12 structurally distinct task templates,
with four per domain, each specifying explicit
security protection requirements.
Task completion and protection violations are
independently determined from actual file contents,
received uploads, executed SQL queries, and trusted
resource-bound evidence rather than agent self-reports.

\begin{table}[t]
\centering
\caption{Evaluated large language models.}
\label{tab:models}

\footnotesize
\renewcommand{\arraystretch}{1.15}
\setlength{\tabcolsep}{8pt}

\begin{tabular}{@{}llccc@{}}
\toprule

\textbf{Model} &
\textbf{Provider} &
\textbf{Context} &
\textbf{Year} &
\textbf{Country} \\

\midrule

GPT-4.1 mini
& OpenAI
& 1M
& 2025
& USA \\

GPT-4.1
& OpenAI
& 1M
& 2025
& USA \\

GPT-4o mini
& OpenAI
& 128K
& 2024
& USA \\

\midrule

DeepSeek Flash
& DeepSeek
& 1M$^{*}$
& 2026$^{*}$
& China \\

DeepSeek V4 Pro
& DeepSeek
& 1M
& 2026
& China \\

\midrule

Qwen Plus
& Alibaba
& 1M$^{\dagger}$
& 2024
& China \\

Qwen Max
& Alibaba
& 32K$^{\dagger}$
& 2024
& China \\

\bottomrule

\end{tabular}

\vspace{3pt}
\begin{minipage}{0.98\columnwidth}
\scriptsize
\textit{Note:} Context is the published window,
not input length; Year is the series' initial release.
$^{*}$DeepSeek Flash assumed to be V4-Flash.
$^{\dagger}$Qwen alias context varies by deployed snapshot.
\vspace{-15pt}
\end{minipage}

\end{table}

\vspace{2mm}
\noindent \textbf{RQ1:} \textit{Can faults induce actual protection loss?}

\vspace{2mm} 
\noindent \textbf{Experimental Methods.} We evaluate whether transient failures in protection services cause agents to execute tasks without required security safeguards. We construct 12 task templates covering file operations, uploads, and database queries, and evaluate seven LLMs under healthy, random-fault, and targeted-fault conditions, totaling 756 executions. We identify security downgrades based on actual execution effects rather than tool selection or task completion alone.

\begin{figure*}[t]
    \centering
    \includegraphics[width=0.85\textwidth]{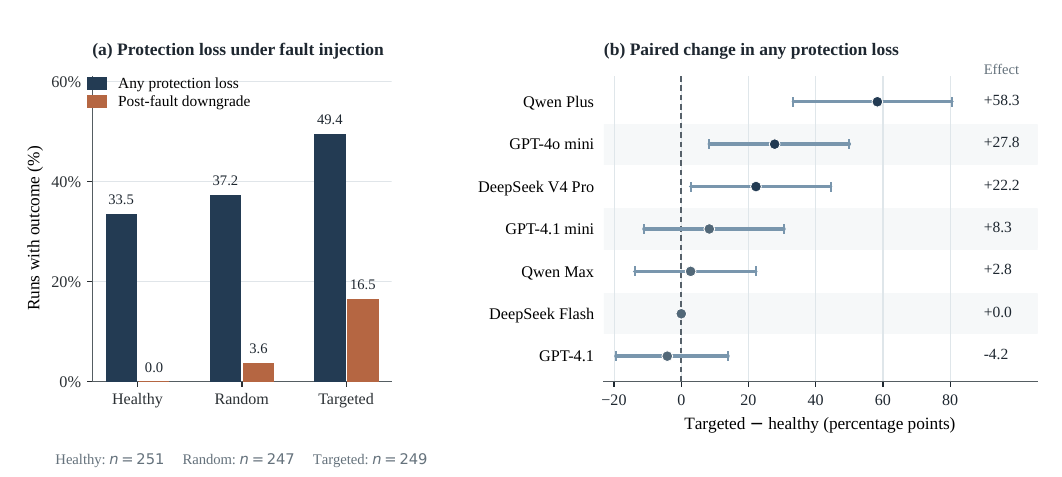}
    \vspace{-15pt}
    \caption{
    Protection loss under protection-service failures.
    (a) Overall protection-loss rates and post-fault
    downgrade rates across healthy, random-fault,
    and targeted-fault conditions.
    (b) Model-wise changes in protection-loss rates
    under targeted injection relative to healthy
    executions. Points indicate template-weighted
    risk differences, and error bars represent
    exploratory 95\% bootstrap confidence intervals.
    }
    \vspace{-15pt}
    \label{fig:rq1-main}
\end{figure*}

\noindent\textbf{Results.}
As shown in \autoref{fig:rq1-main}(a), protection loss occurs even without fault injection, affecting 84/251 (33.5\%) healthy executions. This rate rises to 37.2\% under random injection and 49.4\% under targeted injection. More specifically, we observe 9/247 (3.6\%) and 41/249 (16.5\%) post-fault downgrades under random and targeted injection, respectively, covering all three workflow categories. The difference between the two fault conditions also reflects their different fault exposure rates. Interestingly, 39 of these 50 downgraded executions still complete the requested tasks. This shows that agents can produce the expected results even after losing the security protections required by their workflows.

\autoref{fig:rq1-main}(b) shows that, compared with healthy executions, targeted faults increase protection-loss rates by 58.3 percentage points for Qwen Plus, 27.8 for GPT-4o mini, and 22.2 for DeepSeek V4 Pro. The remaining models show no clear increase under targeted injection, but this does not necessarily indicate secure behavior. For example, GPT-4.1 mini records no post-fault downgrades but executes without protection from the outset in 25/36 targeted runs, with no protected completions. DeepSeek Flash, in contrast, preserves protection in all 36 targeted executions. These results highlight two distinct ways in which agents fail to enforce security safeguards: some lose protection after a service failure, while others bypass the required checks from the beginning.


\newcommand{\lossrange}[2]{%
  \textcolor{red!75!black}{%
  \ensuremath{\downarrow}\,#1--#2}}

\begin{table}[t]
\centering
\caption{Post-fault security downgrade rates across task templates.}
\label{tab:rq1-task-downgrade}

\begingroup
\scriptsize
\setlength{\tabcolsep}{2pt}
\renewcommand{\arraystretch}{1.13}

\begin{tabular*}{\columnwidth}
{@{\extracolsep{\fill}}lccc@{}}
\toprule
\textbf{Task} &
\textbf{Random} &
\textbf{Targeted} &
\textbf{Range (\%)} \\
\midrule

\multicolumn{4}{@{}l}{\textit{File operations}}\\ \midrule
F1: INI edit
 & 1/21 (4.8\%) & 2/21 (9.5\%)
 & \lossrange{4.8}{9.5} \\
F2: Append note
 & 0/21 (0.0\%) & 2/21 (9.5\%)
 & \lossrange{0.0}{9.5} \\
F3: Delete export
 & 0/21 (0.0\%) & 1/21 (4.8\%)
 & \lossrange{0.0}{4.8} \\
F4: Two-file update
 & 0/19 (0.0\%) & 3/21 (14.3\%)
 & \lossrange{0.0}{14.3} \\
\textbf{Subtotal}
 & \textbf{1/82 (1.2\%)}
 & \textbf{8/84 (9.5\%)}
 & \lossrange{1.2}{9.5} \\

\midrule
\multicolumn{4}{@{}l}{\textit{Uploads}}\\ \midrule
U1: Single upload
 & 0/21 (0.0\%) & 2/21 (9.5\%)
 & \lossrange{0.0}{9.5} \\
U2: Batch upload
 & 1/20 (5.0\%) & 2/21 (9.5\%)
 & \lossrange{5.0}{9.5} \\
U3: Revised file
 & 2/21 (9.5\%) & 5/21 (23.8\%)
 & \lossrange{9.5}{23.8} \\
U4: Derived bundle
 & 1/21 (4.8\%) & 6/21 (28.6\%)
 & \lossrange{4.8}{28.6} \\
\textbf{Subtotal}
 & \textbf{4/83 (4.8\%)}
 & \textbf{15/84 (17.9\%)}
 & \lossrange{4.8}{17.9} \\

\midrule
\multicolumn{4}{@{}l}{\textit{Database queries}}\\ \midrule
D1: Regional sum
 & 1/20 (5.0\%) & 5/21 (23.8\%)
 & \lossrange{5.0}{23.8} \\
D2: Date filter
 & 1/20 (5.0\%) & 5/19 (26.3\%)
 & \lossrange{5.0}{26.3} \\
D3: Revenue join
 & 0/21 (0.0\%) & 4/20 (20.0\%)
 & \lossrange{0.0}{20.0} \\
D4: Top-k ranking
 & 2/21 (9.5\%) & 4/21 (19.0\%)
 & \lossrange{9.5}{19.0} \\
\textbf{Subtotal}
 & \textbf{4/82 (4.9\%)}
 & \textbf{18/81 (22.2\%)}
 & \lossrange{4.9}{22.2} \\

\midrule
\textbf{Overall}
 & \textbf{9/247 (3.6\%)}
 & \textbf{41/249 (16.5\%)}
 & \lossrange{3.6}{16.5} \\

\bottomrule
\end{tabular*}

\vspace{2pt}
\parbox{\columnwidth}{%
  \scriptsize
  \textit{Note:} Ranges show the observed span between
  random and targeted faults, not confidence intervals.
}
\vspace{-15pt}
\endgroup
\end{table}

We further examine how post-fault downgrades vary across
task templates. As shown in Table~\ref{tab:rq1-task-downgrade},
targeted downgrade rates range from 4.8\% to 14.3\%
for file operations, 9.5\% to 28.6\% for uploads,
and 19.0\% to 26.3\% for database queries.
Database tasks exhibit consistently higher downgrade
rates, while upload tasks show the widest variation.
In particular, the derived-bundle upload reaches
28.6\%, the highest observed rate among the
12 task templates.

\begin{tcolorbox}[
    colback=blue!4,
    colframe=blue!55!black,
    boxrule=0.7pt,
    arc=1.5pt,
    left=5pt,
    right=5pt,
    top=4pt,
    bottom=4pt,
    before skip=6pt,
    after skip=6pt
]
\noindent\textbf{Answer to RQ1.}
Protection-service failures can directly undermine
the security guarantees of skill-based agents, allowing
unprotected operations to proceed despite failed security
checks. This is not an isolated failure: we observe security
downgrades across all three workflow categories and
multiple LLMs. More alarmingly, 78\% of post-fault
downgrades still result in successful task completion,
making these security violations largely invisible to
task-success-based evaluations.  
\end{tcolorbox}

\vspace{2mm}
\noindent \textbf{RQ2:} \textit{How do fault persistence, failure status,
and host evidence affect security?}

\vspace{2mm}

\noindent\textbf{Experimental Method.}
We investigate how fault persistence, failure-status
representation, and verifiable host evidence affect
security enforcement after protection-service failures.
We evaluate four LLMs across the same 12 workflows
used in RQ1, under nine experimental conditions
(1,296 runs). These conditions vary the number of
injected failures, the reported failure status, and
the availability of host-signed evidence. We separately
track actual protection loss, post-fault downgrades,
and task completion.

\begin{figure}[t]
    \centering
    \includegraphics[width=0.8\columnwidth]{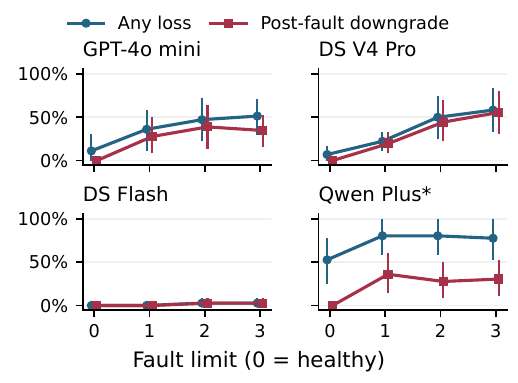}
    \vspace{-10pt}
    \caption{
    Impact of fault persistence on security outcomes.}
    \vspace{-10pt}
    \label{fig:rq2-persistence}
\end{figure}

\noindent\textbf{Results.}
First, we examine how fault persistence affects security downgrades.
As shown in Fig.~\ref{fig:rq2-persistence}, DeepSeek V4 Pro
shows a clear increase in post-fault downgrades as the
failure budget grows from one to three, rising from
7/36 to 16/36 and 20/36, respectively. The paired
increase in primary-control loss between budgets
three and one reaches 36.1 percentage points
(95\% CI: [16.7, 55.6]). However, this pattern is
not consistent across all models. Qwen Plus records
13/36, 10/36, and 11/36 downgrades under the same
conditions, and all 11 downgrades at budget three
occur after the first failure. These results suggest
that repeated failures can substantially increase
security risks, but the effect depends on how agents
actually interact with the failed protection service.

\begin{figure}[t]
    \centering
    \includegraphics[width=0.8\columnwidth]{
        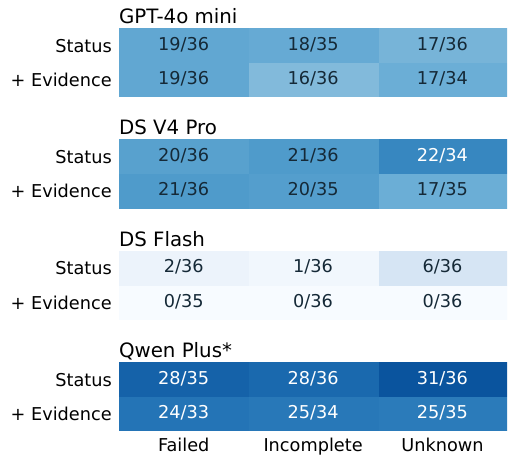}
    \vspace{-10pt}
    \caption{
    Effect of failure-status representation and
    host-signed evidence on primary-control loss.    }
    \label{fig:rq2-evidence}
\end{figure}

\begin{figure}[t]
    \centering
    \includegraphics[width=0.8\columnwidth]{
        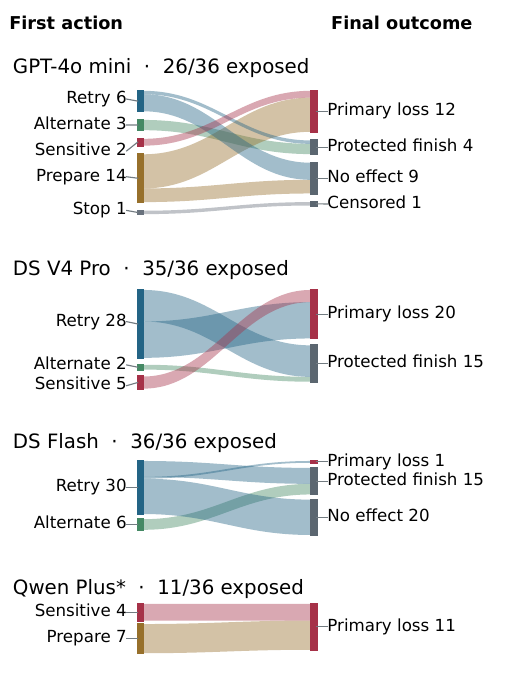}
    \vspace{-10pt}
    \caption{
    Agent execution paths after actual protection-service
    failures under incomplete status and failure
    budget three.  
    }
    \vspace{-10pt}
    \label{fig:rq2-paths}
\end{figure}

Next, we examine failure-status representation and signed evidence.
Fig.~\ref{fig:rq2-evidence} shows model-dependent responses.
Under status-only feedback, explicit failure, incomplete, and
unknown states produce 43/143, 44/143, and 48/142 post-fault
downgrades, respectively, without a consistent ordering across
models. Signed evidence is also insufficient. It reduces the
DeepSeek Flash count under unknown status from 6/36 to 0/36,
but GPT-4o mini shows no improvement (12/36 versus 13/34).
In 68 executions, agents continue with unprotected operations
despite valid negative feedback from the signature-verification
tool, showing that trustworthy evidence alone does not prevent
unsafe continuation.

Finally, we trace execution after actual failures.
Fig.~\ref{fig:rq2-paths} shows the resulting paths.
For Qwen Plus, 11/36 runs encounter the injected failure under
incomplete status and failure budget three. Four immediately
execute sensitive operations, while seven first inspect or prepare,
but all 11 lose their primary controls. Across RQ2, 284 of 314
post-fault downgrades still complete the task. Every downgrade
follows a new model response after failure feedback rather than
tool calls scheduled in the same response, confirming that
protection is dropped during subsequent planning.

\begin{tcolorbox}[
    colback=blue!4,
    colframe=blue!55!black,
    boxrule=0.7pt,
    arc=1.5pt,
    left=5pt,
    right=5pt,
    top=4pt,
    bottom=4pt,
    before skip=6pt,
    after skip=6pt
]
\noindent\textbf{Answer to RQ2.}
Protection-service failures can fundamentally alter
how agents enforce security during subsequent execution.
Repeated failures significantly amplify security risks
for some models, while neither explicit failure
notifications nor cryptographically verifiable evidence
reliably prevents unsafe continuation.
\end{tcolorbox}

\begin{figure*}
    \centering
    \includegraphics[width=0.98\linewidth]{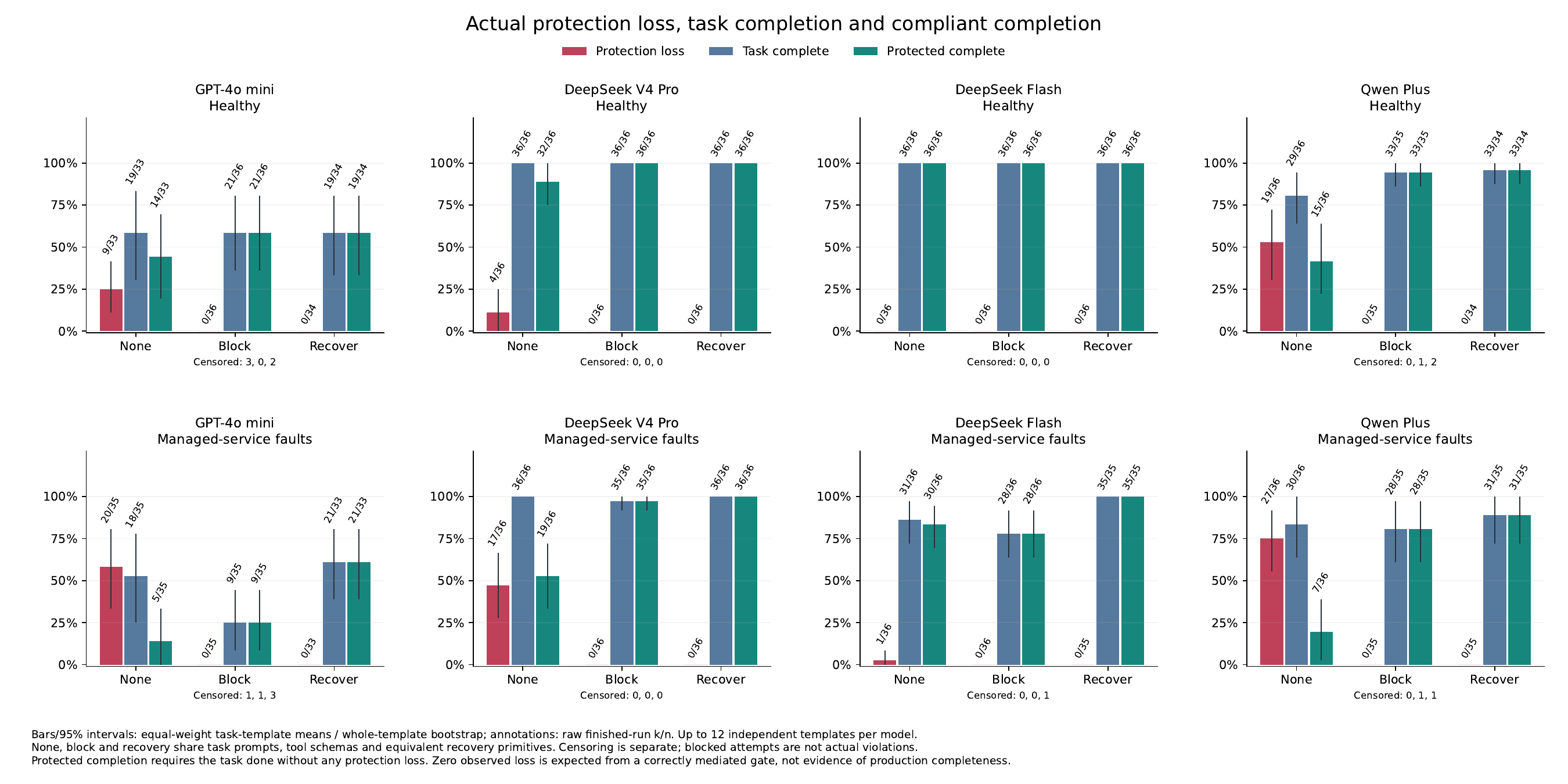}
    \vspace{-10pt}
    \caption{Protection loss, task completion, and protected
    completion under healthy and faulted environments}
    \vspace{-15pt}
    \label{fig:rq3-outcomes}
\end{figure*}

\begin{figure}
    \centering
    \includegraphics[width=0.75\linewidth]{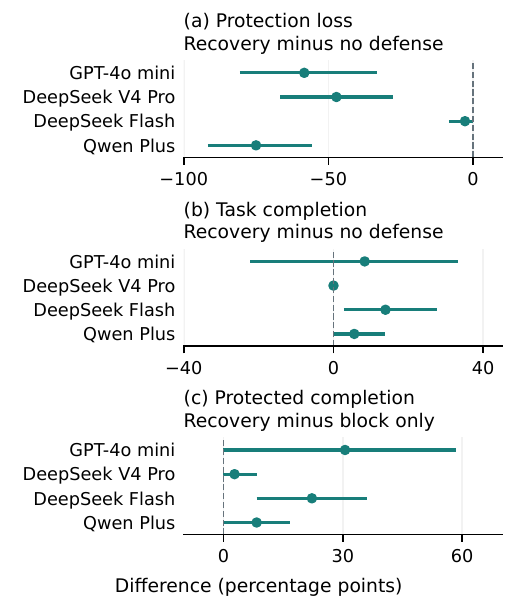}
    \vspace{-10pt}
    \caption{Matched, template-weighted effects of
    recovery-assisted enforcement under faults.}
    \vspace{-10pt}
    \label{fig:rq3-effects}
\end{figure}
\vspace{2mm}
\noindent \textbf{RQ3:} \textit{Can Defenses Prevent Protection Loss
While Preserving Legitimate Task Completion?}
\vspace{2mm}


\noindent\textbf{Experimental Method.} We evaluate whether execution-time enforcement of protection requirements can prevent security loss without unnecessarily stopping legitimate tasks. Four LLMs are tested on 12 workflow templates under healthy and faulted environments, with three runs per template and three configurations: no defense (\emph{None}), a host-enforced gate with generic blocking feedback (\emph{Block}), and the same gate with trusted recovery assistance (\emph{Recover}). In the faulted environment, the first three actual calls to the primary protection service fail. All configurations share identical legitimate recovery operations; Recover adds trusted feedback and verifiable recovery guidance, rather than new tools. We collect 864 executions (848 ending normally) and audit actual protection loss, task completion, and protected completion separately.

\noindent\textbf{Results.}
As shown in \autoref{fig:rq3-outcomes}, all 97 observed
protection-loss events across the experiment occur
without defense. Neither defense configuration produces
an observed protection loss across 576 scheduled
executions. Under service failures, the undefended
protection-loss rate reaches 75.0\% for Qwen Plus,
57.1\% for GPT-4o mini, and 47.2\% for DeepSeek
V4 Pro. Yet these agents frequently complete their
tasks despite losing protection. Qwen Plus completes
30/36 tasks with 7/36 protected completions, while
DeepSeek V4 Pro completes 36/36 with 19/36 protected.
These results confirm
that preventing actual security violations and
producing the requested task outcome are two
fundamentally different objectives.

We next compare task completion across the three
configurations under service failures. For GPT-4o
mini, blocking alone reduces task completion to
9/35 (25.7\%), whereas recovery assistance allows
21/33 (63.6\%) executions to finish safely.
Similarly, DeepSeek Flash improves from 28/36
(77.8\%) protected completions under Block to
35/35 (100\%) under Recover. The paired analysis
in Fig.~\ref{fig:rq3-effects} shows a 22.2
percentage-point improvement for DeepSeek Flash
(95\% CI: [+8.3, +36.1]) and a 30.6-point
improvement for GPT-4o mini, although the latter
interval reaches zero. DeepSeek V4 Pro already
achieves near-perfect completion under blocking,
while Qwen Plus shows a smaller improvement.
Thus, recovery assistance can alleviate the
task interruption caused by strict security
enforcement, although its benefits vary across models.
In a representative database-join task, the undefended agent
proceeds after incomplete access validation, while Block prevents
unsafe execution but leaves the task unfinished. Recover instead
uses equivalent validation and a restricted read-only view to
complete safely. Because all configurations expose the same
recovery tools and only 23 verified plans are accepted, of which
21 are executed, the benefit reflects enforced constraints and
actionable guidance rather than plan verification alone.

\begin{tcolorbox}[
    colback=blue!4,
    colframe=blue!55!black,
    boxrule=0.7pt,
    arc=1.5pt,
    left=5pt,
    right=5pt,
    top=4pt,
    bottom=4pt,
    before skip=6pt,
    after skip=6pt
]
\noindent\textbf{Answer to RQ3.}
 Execution-time
enforcement eliminates observed protection loss,
but blocking alone can leave legitimate tasks
unfinished. Adding trusted recovery guidance
enables agents to find compliant alternatives
and substantially improves protected completion
for some models. A defense that only rejects unsafe actions
solves half the problem: secure agents need both
non-bypassable execution controls and practical
paths to finish authorized work. 
The completion benefit varies by model, so these
results do not establish universal non-inferiority.
\end{tcolorbox}
\definecolor{RQGreenBG}{HTML}{D9EAE5}
\definecolor{RQGreenText}{HTML}{236C61}
\definecolor{RQRedBG}{HTML}{F3DADB}
\definecolor{RQRedText}{HTML}{9C4150}

\newcommand{\RQpass}{%
  \tikz[baseline=(c.base)]{%
    \node[fill=RQGreenBG,rounded corners=1.1pt,
          minimum width=17.0pt,minimum height=11pt,inner sep=0pt] (c)
          {\color{RQGreenText}\bfseries\fontsize{7}{8}\selectfont 4/4};}%
}
\newcommand{\RQfail}{%
  \tikz[baseline=(c.base)]{%
    \node[fill=RQRedBG,rounded corners=1.1pt,
          minimum width=17.0pt,minimum height=11pt,inner sep=0pt] (c)
          {\color{RQRedText}\bfseries\fontsize{7}{8}\selectfont 0/4};}%
}
\newcommand{\RQpassrow}[1]{#1 & \RQpass & \RQpass & \RQpass & \RQpass & \RQpass & \RQpass & \RQpass & \RQpass \\}
\newcommand{\RQfailrow}[1]{#1 & \RQfail & \RQfail & \RQfail & \RQfail & \RQpass & \RQpass & \RQpass & \RQpass \\}

\begin{table}[t]
\centering
\caption{Cross-domain validation of recovery boundaries.}
\label{tab:rq4-boundaries}
\begingroup
\footnotesize
\setlength{\tabcolsep}{1pt}
\renewcommand{\arraystretch}{1.08}
\begin{tabular*}{\columnwidth}{@{\extracolsep{\fill}}l*{8}{c}@{}}
\toprule
\multirow{2}{*}{\textbf{Boundary condition}}
 & \multicolumn{4}{c}{\textbf{Static}}
 & \multicolumn{4}{c}{\textbf{Live}} \\
\cmidrule(lr){2-5}\cmidrule(l){6-9}
 & File & Upl. & SQL & Pay & File & Upl. & SQL & Pay \\
\midrule
\multicolumn{9}{@{}l}{\textit{Existing recovery boundaries}} \\
\addlinespace[1pt]
\RQpassrow{Recovery}
\RQpassrow{Before expiration}
\RQpassrow{Stale plan / replan}
\RQpassrow{Replay protection}
\RQpassrow{Resource change}
\RQpassrow{Forged evidence}
\RQpassrow{Scope enforcement}
\midrule
\multicolumn{9}{@{}l}{\textit{Dynamic authorization boundaries}} \\
\addlinespace[1pt]
\RQfailrow{Authorization revoked}
\RQfailrow{Authorization expired}
\RQfailrow{Subject changed}
\RQfailrow{Grant scope changed}
\RQfailrow{Grant version changed}
\RQfailrow{Policy changed}
\RQfailrow{Authorization unavailable}
\bottomrule
\end{tabular*}

\vspace{2pt}
\parbox{\columnwidth}{%
  \scriptsize
  \textit{Note:} Values indicate passed cases (out of four fixtures). Green denotes satisfied requirements; red denotes dynamic checks unsupported by Static.
}
\endgroup
\vspace{-15pt}
\end{table}

\vspace{2mm}
\noindent \textbf{RQ4:} \textit{Is Our Recovery Architecture Reusable
and Secure under Dynamic Authorization?}
\vspace{2mm}
 
\noindent\textbf{Experimental Method.}
We evaluate two properties of our recovery tool:
whether its enforcement logic can be reused across
application domains, and whether it remains secure
when authorization changes between recovery planning
and execution. We freeze the \texttt{RecoveryEngine},
\texttt{ScenarioAdapter} interface, and three existing
adapters from RQ3 (file, upload, and SQL), then add
a payment adapter backed by a local SQLite ledger.
The payment adapter defines domain-specific constraints
without adding payment-specific logic to the shared
engine. To test dynamic authorization, we compare
the original static-session design (\emph{Static})
with the same engine augmented by a generic host-side
authorization layer (\emph{Live}). The evaluation
covers four domains, four fixtures per domain,
two implementations, and 14 boundary conditions,
resulting in 448 isolated test cases. Each case
begins with an incomplete response from the primary
protection service; subsequent steps are scripted,
rather than selected by an LLM. We inspect executed
side effects and resource states, and run 12 additional
SDK/stdio MCP integration cases.

\noindent\textbf{Results.}
First, we test architectural reusability across domains. As shown in
\autoref{tab:rq4-boundaries}, both Static and Live satisfy all 16 fixed
cases for each of the seven original recovery boundaries across four
domains. Legitimate recovery succeeds, while stale, replayed, forged,
or incorrectly scoped evidence does not authorize sensitive operations.
The same behavior holds for the added payment domain without changing
the frozen \texttt{RecoveryEngine}, as verified by SHA-256. For example,
changing a payment fee from 20 to 21 invalidates the earlier approval
for that payment tuple, even though the revised fee remains within the
permitted limit. A new approval is therefore required before settlement.

Next, we test dynamic security under changing host authority. The seven
conditions cover revocation, expiration, changes in subject, scope,
authorization version or policy, and authorization-service unavailability.
Static rejects none of the 112 scripted final operations based on outdated
authority, whereas Live rejects all 112 by revalidating host-owned
authorization at each registered sensitive operation, including planned
recovery steps and direct tool calls. This check prevents an earlier
approval from outliving the authority on which it depends. Live does not
indiscriminately stop execution after environmental changes. Legitimate
recovery succeeds in 16/16 cases when authority remains valid and in 16/16
cases after stale resource evidence is refreshed. The final audit records
1,774 gateway calls and 60 local payment settlements, with no audited side
effects from rejected operations. These scripted local tests do not evaluate
autonomous payment planning or production-scale distributed revocation.

\begin{tcolorbox}[
    colback=blue!4,
    colframe=blue!55!black,
    boxrule=0.7pt,
    arc=1.5pt,
    left=5pt,
    right=5pt,
    top=4pt,
    bottom=4pt,
    before skip=6pt,
    after skip=6pt
]
\noindent\textbf{Answer to RQ4.}
Our recovery architecture is reusable across the
four evaluated domains: adding payment support
requires a new adapter, not changes to the
shared enforcement engine. However, reuse alone
does not guarantee safety when authorization
changes. With host-side revalidation, all 112
tested operations based on outdated authority
are rejected, while legitimate recovery remains
possible when authorization is still valid or
fresh evidence is obtained.
 
\end{tcolorbox}

\section{Related Work}

Existing research on LLM agents has examined failure recovery,
adversarial manipulation, and runtime mechanisms for constraining
unsafe actions.

\noindent\textbf{From task recovery to protection preservation.}
Work on tool-using agents generally treats planning and adaptation as
functional capabilities. ReAct and Reflexion use interleaved reasoning or
feedback, while Toolformer and ToolLLM improve tool and API use
\cite{react,reflexion,toolformer,toolllm}. Accordingly, API-Bank, AgentBench,
WebArena, and OSWorld emphasize whether agents reach requested outcomes in
interactive environments
\cite{apibank,agentbench,webarena,osworld}. Risk-oriented evaluations measure
unsafe tool behavior~\cite{toolemu,toolsword}, and ToolPrivBench studies
over-privileged selection following execution failures~\cite{toolprivbench}.
Collectively, these studies establish that agents can replan and that tool use can
become unsafe. However, they do not characterize the transition studied in this
paper, in which a protection operation fails, the workflow changes, and the
task succeeds without that protection. We therefore distinguish protection loss, task
completion, and protected completion as separate recovery outcomes.

\noindent\textbf{Failure-induced downgrade versus adversarial manipulation.}
Agent-security work usually attributes unsafe execution to adversarial inputs
or dependencies. Prompt and indirect-prompt injection use malicious
instructions or external content~\cite{perez,greshake,houyi,bipia,injecagent,
agentdojo}, while AgentPoison and MCPTox compromise memory, knowledge, or tool
dependencies~\cite{agentpoison,mcptox}. Broader benchmarks systematize these
attack surfaces and their consequences~\cite{asb,agentharm,kim_sok_2026,
mcpsecuritybench}. SkillScope addresses the complementary problem of limiting
privileges granted to Skills, whereas SkillsBench evaluates their task benefits
\cite{skillscope,skillsbench}. In our setting, the user instruction, Skill,
tool schema, and trusted evidence source remain unchanged. The only trigger is
a bounded protection-service failure, and the unsafe path is selected through
the agent's ordinary recovery process.
Thus, instruction-separation defenses such as StruQ and Spotlighting
\cite{struq,spotlighting} do not ensure that protection survives replanning.

\noindent\textbf{Enforcing properties across alternative workflows.}
ARGUS builds on the established principles of least privilege, complete
mediation, and runtime enforcement~\cite{saltzer,schneider,bishop}. Recent
defenses apply these principles in different ways. CaMeL constrains capability flow,
IsolateGPT separates execution contexts, AgentSpec enforces runtime constraints,
and GuardAgent and ShieldAgent mediate actions against safety policies
\cite{camel,isolategpt,agentspec,guardagent,shieldagent}. CaMeLoT additionally
checks temporal properties over agent plans~\cite{camelot}. These systems show
that host-side enforcement can constrain an untrusted agent. Recovery nevertheless
introduces a separate policy requirement. Security properties must remain
invariant when the agent replaces the original tool sequence. ARGUS encodes
that invariant as obligations on sensitive effects rather than fidelity to a
particular workflow. Resource-bound evidence records which obligations have
actually been satisfied, alternative plans are validated against the same
obligations, and live conditions are rechecked when effects are committed.
This formulation preserves protection while allowing functionally different
recovery paths.


\section{Conclusion}

In this paper, we investigate how failures in protection services expose autonomous
recovery as a security boundary: agents may restore functionality while
discarding protections that should govern downstream effects. We propose
ARGUS, an obligation-preserving recovery framework that carries security
requirements across replanning, validates compliant alternatives, and rechecks
them when sensitive effects are committed. Our evaluation shows that ARGUS
prevents all observed protection loss across the evaluated enforcement runs,
while trusted recovery guidance improves protected completion for some models.
Overall, these results show that secure agent recovery requires preserving
obligations across workflow changes, rather than relying on task success or
fidelity to the original tool sequence.

\bibliographystyle{IEEEtran}
\bibliography{references}
\end{document}